\documentstyle[aps,prb,eqsecnum,epsf,floats]{revtex}
\begin{document}
\title{A metal-insulator transition as a quantum glass problem}
\author{T.R. Kirkpatrick}
\address{Institute for Physical Science and Technology and Department of Physics\\
University of Maryland\\
College Park, Maryland 20742}
\author{D. Belitz}
\address{Department of Physics and Materials Science Institute\\
University of Oregon\\
Eugene, Oregon 97403}
\maketitle

\begin{abstract}
We discuss a recent mapping of the Anderson-Mott metal-insulator transition
onto a random field magnet problem. The most important new idea introduced
is to describe the metal-insulator transition in terms of an order parameter
expansion rather than in terms of soft modes via a nonlinear sigma model.
For spatial dimensions $d>d_c^{+}=6$ a mean field theory gives the exact
critical exponents. For $d=6-\varepsilon $ the critical exponents are
identical to those for a random field Ising model. Dangerous irrelevant
quantum fluctuations modify Wegner's scaling law relating the conductivity
exponent to the correlation or localization length exponent. This
invalidates the bound $s\geq 2/3$ for the conductivity exponent $s$ in $d=3$%
. We also argue that activated scaling might be relevant for describing the
AMT in three-dimensional systems.
\end{abstract}

\section{Introduction}

Metal-insulator transitions of purely electronic origin, i.e. those for
which the structure of the ionic background does not play a role, are
commonly divided into two categories. In one category the transition is
triggered by electronic correlations, or interactions, and in the other it
is driven by disorder. The first case is known as a Mott 
transition,\cite{Mott} and the second one as an Anderson transition.\cite{LR} 
It is believed
that for many real metal-insulator transitions both correlations and
disorder are relevant. The resulting quantum phase transition, which carries
aspects of both types of transitions, we call an Anderson-Mott transition
(AMT).\cite{R}

Until very recently virtually all approaches\cite{R} studied the AMT only in
the vicinity of two dimensions by generalizing\cite{F1} Wegner's theory\cite
{Wegner79} for the Anderson transition. Renormalization-group methods lead
to a critical fixed point in $d=2+\varepsilon $ dimensions, and standard
critical behavior with power-law scaling was found. However, the framework
of these theories does not allow for an order parameter (OP) description of
the AMT, and does not lead to a simple Landau or mean-field theory.\cite{HL}
As a result, the physics driving the AMT remains relatively obscure in this
approach, compared to standard theories for other phase transitions. An
alternative line of attack has recently been explored by the present 
authors.\cite{Letter1,Letter2,ZPhys,belitz52}
We have shown that an OP description of the AMT is possible with the
tunneling density of states (DOS) as the OP. A simple Landau theory then
yields the exact critical exponents, above the upper critical dimension, $%
d_c^{+}=6$. In this respect the AMT is conceptually simpler than the
Anderson transition, which has no known simple OP description, and whose upper
critical dimension may be infinite.

One of the most far-reaching implications of our approach is that the AMT is
in some respects similar to magnetic transitions in random fields.
Qualitatively, this can be understood as follows. Consider a model of an
interacting disordered electron gas. In terms of anticommuting Grassmann
fields, $\bar{\psi}$ and $\psi $, the action can be written,\cite{R}

\begin{equation}
S=S_{{\rm k}}+S_{{\rm dis}}+S_{{\rm int}}\quad ,  \label{eq:1.1}
\end{equation}

\noindent with,

\begin{mathletters}
\label{eqs:1.2}
\begin{equation}
S_{{\rm k}} = -\sum_{\sigma} \int dx\ \bar{\psi}_\sigma (x)\,\left[
\partial_{\tau} -\frac{\nabla ^2}{2m} - \mu\right]\,\psi_{\sigma}(x)\quad,
\label{eq:1.2a}
\end{equation}

\noindent 
the kinetic or free part of S,

\begin{equation}
S_{{\rm dis}}=-\sum_{\sigma} \int dx\ u({\bf x})\,\bar{\psi}_{\sigma}(x)
\,\psi_{\sigma}(x)\quad ,  \label{eq:1.2b}
\end{equation}

\noindent 
the disorder part of S, and

\begin{equation}
S_{{\rm int}}=-\frac{\Gamma }{2}\,\sum_{\sigma _1,\sigma _2}\int dx\ \bar{%
\psi}_{\sigma _1}(x)\,\bar{\psi}_{\sigma _2}(x)\,\psi _{\sigma _2}(x)\, \psi
_{\sigma _1}(x)\quad,  \label{eq:1.2c}
\end{equation}

\noindent 
denoting the interaction part of S. In these equations, $x=({\bf x},\tau )$
with $\tau $ denoting imaginary time, $\int dx\equiv \int d{\bf x}%
\int\limits_0^{1/T}d\tau $, $m$ is the electron mass, $\mu $ is the chemical
potential, $\sigma $ is a spin label, and for simplicity we have assumed an
instantaneous point-like electron-electron interaction with strength $\Gamma$%
. $u({\bf x})$ is a random potential which represents the disorder. For
simplicity we also assume $u({\bf x})$ to be $\delta $-correlated, and to
obey a Gaussian distribution with second moment

\end{mathletters}
\begin{equation}
\{u({\bf x})u({\bf y})\}=\frac {1}{2\pi N_F\tau _{{\rm el}}}\ \delta ({\bf x}%
-{\bf y})\quad,  \label{eq:1.3}
\end{equation}

\noindent 
where the braces denote the disorder average, $N_F$ is the bare DOS per spin
at the Fermi energy, and $\tau_{{\rm el}}$ is the bare elastic mean-free
time. For future use we write $S_{{\rm dis}}$ as,

\begin{equation}
S_{{\rm dis}} = -\sum_{n,\sigma }\int d{\bf x}\ u({\bf x}) \ \bar{\psi}%
_{\sigma ,n}({\bf x})\, \psi _{\sigma ,n}({\bf x})\quad,  \label{eq:1.4}
\end{equation}

\noindent 
where a Matsubara frequency decomposition of $\bar{\psi}(\tau )$ and $\psi
(\tau )$ has been used.

As mentioned above, the most obvious candidate for an OP for the AMT is the
single particle DOS, $N$, at the Fermi level. In terms of Grassmann
variables this quantity is proportional to the zero-frequency limit of the
expectation value of the variable $\bar{\psi}\psi $:

\begin{mathletters}
\label{eqs:1.5}
\begin{equation}
N = {\rm Im}\,N(i\omega_n\rightarrow 0+i0) \quad,  \label{eq:1.5a}
\end{equation}

\noindent 
with,

\begin{equation}
N(i\omega_n) = \frac{-1}{2\pi N_F}\sum_{\sigma}\ \langle\bar\psi_{\sigma ,n}
({\bf x})\,\psi_{\sigma ,n}({\bf x})\rangle  \label{eq:1.5b}
\end{equation}

\noindent 
where we have normalized the DOS by $2N_F$, and the brackets denote an
expectation value with respect to the action $S$.
Note that the so defined DOS is
actually a local DOS, i.e. it depends on {\bf x}. Examining Eqs.\ (\ref
{eq:1.4}) and (\ref{eqs:1.5}), we see that the local OP for the AMT couples
linearly to the random potential, and depending on the sign of $u({\bf x})$
it will favor either an increasing or a decreasing DOS. Similarly, $S_{{\rm %
int}} \sim -\Gamma\, N^2$, i.e. $S_{{\rm int}}$ always favors a decreasing
DOS. We conclude that the interaction term in general frustrates the
disorder term, just like in a random field (RF) magnet problem.

This conclusion has a number of important implications. For example, if
conventional scaling exists at the AMT, then one expects hyperscaling to be
violated due to a dangerous irrelevant variable (DIV), as it is in RF 
magnets.\cite{Grinstein} As a consequence of this, we argue below that Wegner's
scaling law relating the conductivity exponent $s$ to the correlation length
exponent $\nu $ is modified. Furthermore, if the AMT shares all of the
features known to be induced in magnets by a random field, then one would
expect glasslike features and unconventional or activated scaling similar to
what has been predicted\cite{FisherVillain} and observed\cite{BelangerYoung}
in classical RF magnets.

The plan of this paper is as follows. In Section II we give a sketch of our
order-parameter theory of the AMT. An explicit scaling theory near $d=6$ is
constructed. In the first part of Section III we give a general scaling
theory of the AMT, assuming it is a conventional phase transition. In the
second part of this section we review some aspects of an activated scaling
theory for the AMT. We conclude in Section IV with a short discussion.

\section{Formalism and Mean Field Theory}

\subsection{Formalism}

Here we briefly review the formalism we have used to show that at least near 
$d=6$, the AMT and the magnetic transition in a RF Ising model have many 
features in common. For details
we refer to two recent papers.\cite{ZPhys,belitz52}

Our starting point is the nonlinear sigma model (NL$\sigma $M) that has been
used to describe the AMT near two dimensions. The solution procedure we use
near the upper critical dimension is closely analogous to the treatment of
the $O(n)$ symmetric NL$\sigma $M in the limit of large $n$.\cite{ZJ} The NL$%
\sigma $M for the AMT is derived from Eq.\ (\ref{eq:1.1}) by assuming that
all of the relevant physics can be expressed in terms of fluctuations of the
particle number density, the spin density, and the one-particle spectral
density. Technically, this is achieved by making long-wavelength
approximations, and by introducing classical composite operators that are
related to the Grassmannian variables mentioned above. The quenched disorder
is handled by means of the replica trick. The resulting action reads,\cite{R}

\end{mathletters}
\begin{mathletters}
\label{eqs:2.1}
\begin{eqnarray}
S[\tilde{Q}] &=&-\frac 1{2G}\int d{\bf x}\ tr\,\left( {\bf \nabla }\tilde{Q}(%
{\bf x})\right) ^2\ +\ 2H\int d{\bf x}\ tr\,\left( \Omega \tilde{Q}({\bf x}%
)\right)  \nonumber \\
&&-\frac{\pi T}4\sum_{n=s,t}\ \int d{\bf x}\ \left[ \tilde{Q}({\bf x})\gamma
^{(n)}\tilde{Q}({\bf x})\right] \quad ,  \label{eq:2.1a}
\end{eqnarray}
where, 
\begin{eqnarray}
\left[ \tilde{Q}({\bf x})\gamma ^{(s)}\tilde{Q}({\bf x})\right]
&=&K_s\sum\limits_{n_1n_2n_3n_4}\text{ }\delta _{n_1+n_3,n_2+n_4}\text{ }%
\sum\limits_\alpha \text{ }\sum\limits_{r=0,3}\text{ }(-1)^r  \nonumber \\
&&tr\left( (\tau _r\otimes s_0)\,\tilde{Q}_{n_1n_2}^{\alpha \alpha }\right)
\ tr\left( (\tau _r\otimes s_0)\,\tilde{Q}_{n_3n_4}^{\alpha \alpha }\right)
\quad ,  \label{eq:2.1b}
\end{eqnarray}
and 
\begin{eqnarray}
\left[ \tilde{Q}({\bf x})\gamma ^{(t)}\tilde{Q}({\bf x})\right] &=&-K_t\text{ }%
\sum\limits_{n_1n_2n_3n_4}\text{ }\delta _{n_1+n_3,n_2+n_4}\text{ }%
\sum\limits_\alpha \text{ }\sum\limits_{r=0,3}(-1)^r\text{ }
\sum\limits_{i=1}^{3}
\nonumber \\
&&tr\left( (\tau _r\otimes s_i)\,\tilde{Q}_{n_1n_2}^{\alpha \alpha }\right)
\ tr\left( (\tau _r\otimes s_i)\,\tilde{Q}_{n_3n_4}^{\alpha \alpha }\right)
\quad .  \label{eq:2.1c}
\end{eqnarray}
Here $\tilde{Q}$ is a classical field that is, roughly speaking, composed of
two fermionic fields. It carries two Matsubara frequency labels, n and m,
and two replica labels, $\alpha $ and $\beta $. The matrix elements are spin
quaternions, with the quaternion degrees of freedom describing the
particle-hole ($\tilde{Q}\sim \bar{\psi}\psi $) and particle-particle ($%
\tilde{Q}\sim \bar{\psi}\bar{\psi}$) channels, respectively. For simplicity
we restrict ourselves to the particle-hole degrees of freedom. For this case
the matrix elements can be expanded in a restricted spin-quaterion basis,

\end{mathletters}
\begin{equation}
\tilde{Q}_{nm}^{\alpha \beta }=\sum\limits_{r=0,3}\text{ }\sum\limits_{i=0}^{3}%
\text{ }_r^i\tilde{Q}_{nm}^{\alpha \beta }\text{ }(\tau _r\otimes
s_i)\quad,  \label{eq:2.2}
\end{equation}

\noindent with $\tau _{0,1,2,3}$ the quaternion basis, and $s_{0,1,2,3}$ the
spin basis ($s_{1,2,3}=i\sigma _{1,2,3}$ with $\sigma _{1,2,3}$ the Pauli
matrices). The matrix $Q$ is subject to the constraints,\cite{R}
\begin{mathletters}
\begin{equation}
\tilde{Q}^2=1\quad,  \label{eq:2.3a}
\end{equation}

\begin{equation}
tr\,\tilde{Q}=0\quad,  \label{eq:2.3b}
\end{equation}

\begin{equation}
\tilde{Q}^{+}=C^T\tilde{Q}^TC=\tilde{Q}\quad,  \label{eq:2.3c}
\end{equation}
\end{mathletters}%
\noindent 
where $C=i\tau _1\otimes s_2$.

In Eq.\ (\ref{eq:2.1a}), $G=2/\pi \sigma $, with $\sigma $ the bare
conductivity, is a measure of the disorder, and $H=\pi N_F/2$ is a frequency
coupling parameter. $K_s$ and $K_t$ are bare interaction amplitudes in the
spin singlet and spin triplet channels, respectively, and $\Omega
_{nm}^{\alpha \beta } = \delta _{nm}\,\delta _{\alpha \beta }\, \omega
_n\,\tau _0\otimes s_0$, with $\omega _n=2\pi \tau n$, is a bosonic
frequency matrix. Notice that $K_s<0$ for replusive interactions.

The correlation functions of the $\tilde{Q}$ determine the physical quantities.
Correlations of $\tilde{Q}_{nm}$ with $nm<0$ determine the soft
particle-hole modes associated with charge, spin and heat diffusion, while
the DOS is determined by $<\tilde{Q}_{nn}^{\alpha \alpha }>$, i.e., 
$\tilde{Q}_{nm}$
with $nm>0$. It is therefore convenient to separate $\tilde{Q}$ into blocks.

\begin{equation}
\tilde{Q}_{nm}^{\alpha \beta }=\Theta (nm)\,Q_{nm}^{\alpha \beta }+\Theta
(n)\Theta (-m)\,q_{nm}^{\alpha \beta }+\Theta (-n)\Theta (m)\,(q^{\dagger
})_{nm}^{\alpha \beta }\quad .  \label{eq:2.5}
\end{equation}

\noindent 
Normally a NL$\sigma $M is treated by integrating out the massive modes,
i.e., the $Q_{nm}$, to obtain an effective theory for the massless modes,
which are here the diffusion processes described by $q$ and $q^{\dagger}$.
However, since our goal is to obtain a field theory for the OP for the AMT, $%
Q_{nn}$, we instead integrate out the massless q-fields here.

Using standard techniques\cite{ZJ} the above program can be carried out. The
resulting OP field theory for the AMT is,\cite{ZPhys}

\begin{eqnarray}
S[Q] = -\frac 1{2G}\int d{\bf x}\text{ }tr\, \left[\left(\nabla Q({\bf x}%
)\right)^2 + \langle\Lambda\rangle\,\left(Q({\bf x})\right)^2\right] 
\nonumber \\
+ 2H\int d{\bf x}\ tr\,\left(\Omega\, Q({\bf x})\right) + \frac u{2G^2}\int d%
{\bf x}\ tr\,\left[(1-f)\,Q^2({\bf x})\right]  \nonumber \\
- \frac u{4G^2}\int d{\bf x}\ tr\,Q^4({\bf x}) - \frac v{4G^2}\int d{\bf x}\
\left( tr_{+}Q^2({\bf x})\,(tr_{-}Q^2({\bf x})\right)^2+\cdots \quad,
\label{eq:2.6}
\end{eqnarray}

\noindent 
where $tr_{\pm }$ denotes `half-traces' that sum over all replica labels but
only over positive and negative frequencies, respectively: $tr_+ =
\sum_{\alpha}\,\sum_{n\ge 0}$\ , $tr_- = \sum_{\alpha}\,\sum_{n<0}$\ . $f =
f(<\Lambda >)$ is a matrix with elements $_r^if_{nm}^{\alpha \beta }=$ $%
\delta _{ro}\,\delta _{io}\,\delta _{nm}\,f_n$ with f$_n>f_m>0$ for $\vert
n\vert <\vert m\vert $. $f_n$ is an increasing function of disorder, G, and $%
\vert K_s\vert $. $<\Lambda >$ in Eq.\ (\ref{eq:2.6}) is proportional to $%
\Omega /<Q>$, and $u$ and $v$ are finite constants, at least for $d>4$. In
giving Eq.\ (\ref{eq:2.6}) we have neglected terms that can be shown to be
renormalization group (RG) irrelevant near the AMT.

\subsection{Mean-field Theory}

\label{subsec:II.B}

Here we construct a mean-field or saddle-point (SP) solution of Eq.\ (\ref
{eq:2.6}).\cite{MaFisher,ZJ} We look for solutions, $Q_{{\rm sp}%
} $, that are spatially uniform and satisfy,

\begin{equation}
_r^i(Q_{{\rm sp}})_{nm}^{\alpha \beta } = \delta _{ro}\,\delta _{io}\,\delta
_{nm}\,\delta_{\alpha \beta }\, N_n^{(0)}\quad,  \label{eq:2.7}
\end{equation}

\noindent 
where the subscript $(0)$ denotes the SP approximation. The replica,
frequency, and spin-quaternion structures in Eq.\ (\ref{eq:2.7}) are due to
the fact that $<{_r^iQ}_{nm}^{\alpha \beta }>$ has these properties, and
that in the mean-field approximation averages are replaced by the
corresponding SP values.

In the zero-frequency limit, the SP equation of state obtained from Eq.
(2.6) is, 
\begin{mathletters}
\label{eqs:2.8}
\begin{equation}
\left( N_{n=0}^{(0)}\right)^2 = 1 - f_{n=0}(<\Lambda >)=t^{(0)} \quad,
\label{eq:2.8a}
\end{equation}
or 
\begin{equation}
N_{n=0}^{(0)} = \left(t^{(0)}\right)^{1/2}\quad.  \label{eq:2.8b}
\end{equation}
Here $t^{(0)}$ is the mean-field value of the distance from the critical
point, $t$. Equation\ (\ref{eq:2.8b}) yields the mean-field value for the
critical exponent $\beta $,

\end{mathletters}
\begin{equation}
\beta =1/2\quad.  \label{eq:2.9}
\end{equation}
To obtain the remaining mean-field critical exponents we expand $Q$ about
its expectation value, which is proportional to $N$ ,

\begin{equation}
_r^iQ_{nm}^{\alpha \beta }=\delta _{ro}\,\delta _{io}\,\delta_{\alpha
\beta}\, \delta _{nm}\,N_n + \sqrt{2G}\ {_r^i\varphi}_{nm}^{\alpha \beta
}\quad,  \label{eq:2.10}
\end{equation}

\noindent 
where the factor of $\sqrt{2G}$ has been inserted for convenience. The
action $S_{{\rm G}}$ governing Gaussian fluctuations about the mean-field
solution in the critical region then follows from Eq.\ (\ref{eq:2.6}) as,

\begin{mathletters}
\label{eqs:2.11}
\begin{eqnarray}
S_{{\rm G}}[\varphi] = -\int d{\bf x}\ tr\,\left[(\nabla \varphi ({\bf x}%
))^2 + \ell^{(0)}\varphi^2({\bf x})\ + \frac{2u}G\ t^{(0)}\varphi ^2({\bf x}
)\right]  \nonumber \\
- \frac v{2G}\text{ }t^{(0)}\int d{\bf x}\text{ }(tr_{+}\varphi ({\bf x}))%
\text{ }(tr_{-}\varphi ({\bf x}))+O(\varphi ^3)\quad.  \label{eq:2.11a}
\end{eqnarray}
Here 
\begin{equation}
\ell _n^{(0)} = 2GH\omega _n/N_{n\text{ }}^{(0)}\quad,  \label{eq:2.11b}
\end{equation}

\noindent is the SP value of $\Lambda $. All remaining critical exponents
can now be read off Eq.\ (\ref{eq:2.11a}). Comparing the first and third
terms on the r.h.s. yields the correlation length exponent $\nu =1/2.$ With $%
\ell ^{(0)}\sim \omega /Q$, the first and second term gives the dynamical
exponent $z=3.$ Finally, the $\varphi -\varphi $ correlation function near
the transition has the standard Ornstein-Zernike form, which yields the
critical exponents $\gamma =1$ and $\eta =0$. We thus have standard
mean-field values for all static exponents,\cite{MaFisher}

\end{mathletters}
\begin{mathletters}
\label{eqs:2.12}
\begin{equation}
\beta =\nu =1/2 \quad,\quad \gamma =1\quad,\quad \eta =0 \quad,\quad \delta
= 3\quad,  \label{2.12a}
\end{equation}

\noindent and for the dynamical exponent we have,

\begin{equation}
z=3\quad.  \label{eq:2.12b}
\end{equation}

\noindent Inspection of Eq.\ (\ref{eq:2.11a}) shows that the AMT saddle
point is a local minimum and therefore stable.

It is also possible to determine the critical behavior of the transport
coefficients by directly computing the $q-q$ correlation functions and
identifying the change, spin and heat diffusion coefficients $(D_c,D_s,D_h)$%
. Near the mean-field AMT, all three of these coefficients behave in the
same way and vanish like the OP,

\end{mathletters}
\begin{equation}
D_a\sim N_{n=0}^{(0)}/GH\quad ,  \label{eq:2.13}
\end{equation}
with $a=c,s,h$.

The next step in the standard approach for describing any continuous phase
transition is to introduce RG ideas.\cite{MaFisher} In the problem
considered here, application of the RG method accomplishes three things.
First, it generates all additional terms in the action that are consistent
with the symmetry of the problem. Second, it enables us to prove that there
exists an upper critical dimension, $d_c^{+}$, above which mean-field theory
for the critical exponents is exact. Third, it enables us to do an $%
\varepsilon -\exp $ansion below $d_c^{+}$. We begin by noting that Eq.\ (\ref
{eq:2.6}) does not have the RF term we argued for in the Introduction. A
Wilson-type RG procedure generates this term as well as others. It has the
form,

\begin{equation}
S_{RF}=\frac \Delta 2\int d{\bf x}\text{ }\sum\limits_{i=\pm }\text{ }\left(
tr_i\,\varphi ({\bf x})\right)^2\quad.  \label{eq:2.14}
\end{equation}

\noindent In terms of the original fermion action, this contribution arises
from a RF term of the form,

\begin{mathletters}
\label{eqs:2.15}
\begin{equation}
S_{RF}=\sum\limits_{n,\sigma }\int d{\bf x}\text{ }h_n({\bf x})\,\bar{\psi}%
_{\sigma ,n}({\bf x})\,\psi _{\sigma ,n}({\bf x})\quad ,  \label{eq:2.15a}
\end{equation}
where $h_n({\bf x})$ is a random field with 
\begin{equation}
\{h_n({\bf x})\text{ }h_m({\bf x})\}=\theta (nm)\,\frac \Delta {4G}\,\delta (%
{\bf x-y})\quad .  \label{eq:2.15b}
\end{equation}

\noindent All other terms generated by the renormalization process are
irrelevant near the upper critical dimension, $d_c^{+}.$

Standard arguments imply that such a RF term yields $d_c^{+}=6$ instead of
the usual $d_c^{+}=4$.\cite{Grinstein} 
The same arguments also prove that the mean-field
critical behavior quoted above is the exact critical behavior for $%
d>d_c^{+}=6.$ For $d=6-\varepsilon ,$ an $\varepsilon -$expansion of the
critical exponents is possible. The main idea is that under renormalization,
the disorder $\Delta $ scales to infinity while u, the coefficient of the
quartic term in Eq.\ (\ref{eq:2.6}), scales to zero such that their product

\end{mathletters}
\begin{equation}
g = u\,\Delta \quad,  \label{eq:2.16}
\end{equation}

\noindent scales to a stable fixed point value that is of $O(\varepsilon ).$
In all of the other flow equations only the product $g$ appears so that a
stable critical fixed point is obtained. To first order in $\varepsilon
=6-d, $ the resulting critical exponents are,\cite{Letter2,ZPhys}

\begin{mathletters}
\label{eqs:2.17}
\begin{equation}
\nu =\frac 12+\frac \varepsilon {12}\text{ }+O(\varepsilon ^2)\quad,
\label{eq:2.17a}
\end{equation}

\begin{equation}
\gamma =\frac 12-\frac \varepsilon 6\text{ }+O(\varepsilon ^2)\quad,
\label{eq:2.17b}
\end{equation}

\begin{equation}
\delta =3+\varepsilon \text{ }+O(\varepsilon ^2)\quad,  \label{eq:2.17c}
\end{equation}

\begin{equation}
\eta =0+O(\varepsilon ^2)\quad,  \label{eq:2.17d}
\end{equation}

\begin{equation}
z=3-\frac \varepsilon 2+O(\varepsilon ^2)\quad,  \label{eq:2.17e}
\end{equation}

We finally mention that the RG flow properties, $\Delta \rightarrow \infty
,u\rightarrow 0,$ $g\sim O(\varepsilon )$, have an interesting physical
interpretation. $\Delta $ represents the disorder, while $u$ is a measure of
the importance of quantum fluctuations about the SP. Because $u$ determines
physical quantities like the order parameter, this implies that quantum
fluctuations are dangerously irrelevant near the OP driven AMT. This in turn
modifies the standard hyperscaling equalities.

\section{Scaling Descriptions of the Anderson-Mott Transition}

Based on the known or suspected behavior of the random field Ising model,
there are two distinct scaling scenarios one can imagine for the AMT. The
first is a conventional one,\cite{MaFisher} that takes into account in a
general way the dangerous irrelevant variable discussed in Section II. The
second, strikingly different one is new in the context of metal-insulator
transitions, and is called the activated scaling scenario.\cite{FisherVillain}

\subsection{Conventional scaling description of the AMT}

There are standard ways to construct conventional scaling descriptions of
either classical or quantum $(T=0)$ phase transitions. For the random field
like transition considered here, one of the most important features is the
presence of a dangerous irrelevant variable (DIV), namely $u$. Suppose that $%
u$ is characterized by an exponent $\theta$, defined so that $u(b)\sim
b^{-\theta }$. One-loop perturbation theory gives $\theta =2+O(\varepsilon )$%
, but here we keep $\theta $ general. This adds a third independent exponent
to the usual two independent static exponents. In addition, there is the
dynamical scaling exponent $z$. For the case considered here it turns out
that $z$ is not independent, but rather it is equal to the scale dimension
of the field conjugate to the OP. This is due to the fact that RF
fluctuations are much more important than quantum fluctuations. The
dominance of RF fluctuations compared to either thermal or quantum
fluctuations is a general feature of RF problems. The net result, confirmed
explicitly near $d=6$, is,

\end{mathletters}
\begin{equation}
z = y_h = \delta \beta /\nu\quad.  \label{eq:3.1}
\end{equation}

As for the classical RF problem, the DIV $u$, changes $d$ in all scaling
relations to $d-\theta $. For example, near the transition the OP obeys a
scaling or homogeneity relation,

\begin{mathletters}
\label{eqs:3.2}
\begin{equation}
N(t,\Omega ) = b^{-\beta /\nu}\,N(b^{1/\nu}\,t,b^z\,\Omega )\quad,
\label{eq:3.2a}
\end{equation}
with $\beta$ related to $\nu$ and $\eta$ by the usual scaling law, but with $%
d\rightarrow d-\theta $ due to the violation of hyperscaling by the DIV, 
\begin{equation}
\beta =\frac{\nu}{2}\ (d-\theta -2+\eta )\quad.  \label{eq:3.2b}
\end{equation}
Similarly, the exponents $\delta$ and $\gamma$ are given by, 
\end{mathletters}
\begin{mathletters}
\label{eqs:3.3}
\begin{equation}
\delta =(d-\theta +2-\eta )\nu /2\beta\quad,  \label{eq:3.3a}
\end{equation}
\begin{equation}
\gamma =\nu (2-\eta )\quad.  \label{eq:3.3b}
\end{equation}

Next we consider the transport coefficients. The charge, spin, or heat
diffusion coefficients, which we denote collectively by $D$, all scale like
a length squared times a frequency, so that 
\end{mathletters}
\begin{mathletters}
\label{eqs:3.4}
\begin{equation}
D(t,\Omega) = b^{2-z}\,D(t\,b^{1/\nu},\Omega\,b^z) = t^{\nu
(z-2)}\,D(1,\Omega /t^{\nu z})\quad.  \label{eq:3.4a}
\end{equation}
Denoting the static exponent for the diffusion coefficient by $s_D$, defined
by $D(t,\Omega =0)\sim t^{s_D}$, we have found,

\begin{equation}
s_D=\nu (z-2)=\beta -\nu \eta =\frac \nu 2\text{ }(d-2-\theta -\eta )\quad.
\label{eq:3.4b}
\end{equation}

\noindent The behavior of the electrical conductivity $\sigma ,$ which is
related to the charge diffusion coefficient by means of an Einstein
relation, $\sigma =D_c\,\partial n/\partial \mu $, depends on the behavior
of $\partial n/\partial \mu $. If $\partial n/\partial \mu $ has a constant
contribution at the AMT, then $\sigma \sim t^s$ vanishes as $D_c$, so that

\end{mathletters}
\begin{equation}
s=s_D=\frac \nu 2\text{ }(d-2-\theta -\eta)\quad.  \label{eq:3.5}
\end{equation}

\noindent Dimensionally, however, all of the thermodynamic susceptibilities
scale like an inverse volume times a time, which implies a singular part $%
(\partial n/\partial \mu)_s$ of $\partial n/\partial \mu$ that scales like,

\begin{equation}
(\partial n/\partial \mu)_s\,(t,T) = b^{-d+\theta +z}\, (\partial n/\partial
\mu)_s\,(tb^{1/\nu},Tb^z)\quad.  \label{eq:3.6}
\end{equation}

\noindent If there is no constant, analytic, background term, then Eqs.\ (%
\ref{eqs:3.4}) and (\ref{eq:3.6}) give,

\begin{equation}
s=\nu (d-2-\theta )\quad.  \label{eq:3.7}
\end{equation}

\noindent In either case, Wegner's scaling law $s=\nu (d-2)$,\cite{Wegner76}
which previously had been believed to hold for the AMT as well as for the
Anderson transition, is violated, unless Eq. (3.5) holds {\em and} $\theta
=2-d-\eta $. Finally, we note that Eqs.\ (\ref{eq:3.5}) and (\ref{eq:3.7})
are identical if $\eta =\theta +2-d$, and that this result is consistent
with Wegner scaling apart from the replacement $d\rightarrow d-\theta $, and
with $\partial n/\partial \mu $ being noncritical across the AMT. However,
Eq.\ (\ref{eq:3.2b}) shows that this result is not consistent with a
vanishing OP unless the theory has multiple dynamical scaling 
exponents.\cite{R}

\subsection{Activated scaling description of the AMT}

\label{subsec:III.B}

An important characteristic of a glass transition is the occurrence of
extremely long time scales. While critical slowing down at an ordinary
transition means that the critical time scale grows as a power of the
correlation length, $\tau \sim \xi ^z$ with $z$ the dynamical scaling
exponent, at a glass transition the critical time scale grows exponentially
with $\xi ,$

\begin{equation}
\ln (\tau /\tau _0)\sim \xi ^\psi\quad ,  \label{eq:3.8}
\end{equation}

\noindent with $\tau _0$ a microscopic time scale, and $\psi $ a generalized
dynamical scaling exponent. Effectively, Eq.\ (\ref{eq:3.8}) implies $%
z=\infty $. As a result of such extreme slowing down, the system's
equilibrium behavior near the transition becomes inaccessible for all
practical purposes. That is, realizable experimental time scales are not
sufficient to reach equilibrium, and one says that the system falls out of
equilibrium. It has been proposed\cite{FisherVillain} that the phase
transition in classical RF magnets is of this type, and there are
experimental observations that seem to corroborate this suggestion.

Here we speculate that the analogy between RF magnets and the AMT leads to
such `activated' scaling for the AMT as well. For this quantum phase
transition one expects time and inverse temperature to show the same scaling
behavior, irrespective of whether the critical slowing down follows an
ordinary power law, or Eq.\ (\ref{eq:3.8}). Quantum mechanics thus makes it
very difficult to observe the static scaling behavior, since it requires
exponentially small temperatures. Thus from Eq.\ (\ref{eq:3.8}) we see that
static zero temperature scaling will be observed only if

\begin{equation}
T<T_0\exp \left(-\xi ^\psi\right)\quad,  \label{eq:3.9}
\end{equation}

\noindent with $T_0$ some microscopic temperature scale on the order of the
Fermi temperature $\sim T_F$. This is potentially a crucial point in the
interpretation of experimental data.

Activated scaling, as described by Eq.\ (\ref{eq:3.8}), follows from a
barrier picture of the system's free energy landscape. The physical idea we
have in mind is that while a repulsive electron-electron interaction always
leads to a decrease in the local DOS, the random potential can in general
lead to an increase in the local DOS as well. The competition between these
two effects leads to frustration and to, for example, large insulating
clusters within the metallic phase. Delocalizing these large clusters
requires energy barriers to be overcome, which are assumed to grow like $\xi
^\psi $ as the AMT is approached. A further notion of the barrier model is
that the frequency or temperature argument of the scaling function is
expected to be $\ln (\tau/\tau _0)/\ln (T_0/T),$ rather than $\tau\, T$ as
in, for example, Eq.\ (\ref{eqs:3.2}) and (\ref{eqs:3.4}). The reason is
that one expects a very broad distribution of energy barriers. The natural,
self-averaging, variable is therefore $\ln \tau $ rather than $\tau $.

It makes physical sense to assume scaling forms only for self-averaging
quantities. For a system with quenched disorder it is known that the
free energy is self-averaging, while the partition function is not, and
correlation functions in general are not, either. Therefore, all
thermodynamic quantities, which can be obtained as partial derivatives of
the free energy, are self-averaging. For a general thermodynamic quantity, $%
Q $, one therefore expects a homogeneity law\cite{belitz52}

\begin{equation}
Q(t,T)=b^{-x_Q}\,F_Q\left(t\,b^{1/\nu},\frac{b^\psi }{\ln (T_0/T)}%
\right)\quad,  \label{eq:3.10}
\end{equation}

\noindent where $x_Q$ is the scale dimension of $Q,$ and $F_Q$ is a scaling
function. For example, for the DOS one expects,

\begin{equation}
N(t,T) = b^{-\beta /\nu}\,F_N\left(t\,b^{1/\nu},\frac{b^\psi}{\ln (T_0/T)}
\right) \quad,  \label{eq:3.11}
\end{equation}

\noindent with $\beta $ still given by Eq.\ (\ref{eq:3.2b}). Alternatively,
Eq.\ (\ref{eq:3.11}) can be written,

\begin{equation}
N(t,T) = \frac 1{\left[\ln (T_0/T)\right]^{\beta /\nu \psi }}\
G_N\left[t^{\nu \psi }\ln(T_0/T)\right]\quad.  \label{eq:3.12}
\end{equation}

\noindent The scaling function $G_N$ is related to the function $F_N$ in
Eq.\ (\ref{eq:3.4b}) by $G_N(x)=F_N(x^{1/\nu \psi },1)$, and has the
properties $G_N(x\rightarrow \infty )\sim x^{\beta /\nu \psi }$, and $%
G_N(x\rightarrow 0)\rightarrow {\rm const.}$.

Equation\ (\ref{eq:3.12}) makes a qualitative prediction that can be used to
check experimentally for glassy aspects of the AMT: Measurements of the
tunneling DOS very close to the transition should show an anomalously slow
temperature dependence, i.e., $N$ should vanish as some power of $\ln T$
rather than as a power of $T$. While in principle this should be
straightforward, similar checks for the RF problem have shown that a very
large dynamical or temperature range is needed to produce conclusive results.

Other thermodynamic quantities can be considered and are discussed in detail
elsewhere. One chief result is the occurence of a `Griffiths phase', where
both the spin susceptibility and the specific heat expansion coefficient are
singular away from the AMT.\cite{belitz52}

We conclude this subsection by considering the electrical conductivity. Let $%
\tilde{\sigma}$ be the unaveraged conductivity, and $\sigma_0$ a suitable
conductivity scale, e.g., the Boltzmann conductivity. Since $\tilde{\sigma}$
is directly related to a relaxation time, we expect it not to be
self-averaging, while its logarithm should be self-averaging. We define $%
\ell _{\sigma}=<\log (\sigma _0/\tilde{\sigma})>$ 
and assume it is self-averaging and that it satisfies,

\begin{eqnarray}
\ell_{\sigma}(t,T) = 
b^\psi F_\sigma \left( t\,b^{1/\nu },\frac {b^\psi}{\ln(T_0/T)%
} \right)  \nonumber \\
= \ln (t_0/T)\ G_{\sigma} \left(t^{\nu \psi }\ln (T_0/T)\right)\quad.
\label{eq:3.13}
\end{eqnarray}

\noindent Notice that the scale dimension of $\ell _\sigma $ is necessarily $%
\psi$, since $\psi$ characterizes the free energy barriers near the AMT.

As a measure of the conductivity, let us define, 
\begin{equation}
\sigma (t,T) \equiv \sigma _0\exp (-\ell _\sigma )\quad.  \label{eq:3.14}
\end{equation}

\noindent One can argue on physical grounds that $G_\sigma (x\rightarrow
\infty )\sim 1/x$. This yields,

\begin{mathletters}
\label{eqs:3.15}
\begin{equation}
\sigma (t,T=0) \sim \exp (-1/t^{\nu \psi })\quad,  \label{eq:3.15a}
\end{equation}

\noindent and

\begin{equation}
\sigma (t=0,T) \sim T^{\,G_\sigma (0)}\quad.  \label{eq:3.15b}
\end{equation}

\noindent Note that at zero temperature, $\sigma$ vanishes exponentially
with $t$, and that at the critical point $\sigma$ vanishes like a
nonuniversal power of $T$.

\section{Discussion}

We conclude by briefly summarizing our order parameter description of the
AMT and its relation to the random field magnet problem. We also make a few
additional comments on the experimental situation.

The RF nature of the AMT was made plausible in the introduction. In order to
derive this result it is necessary to have an OP description of the AMT. In
Section\ II we illustrated how to obtain an OP field theory for the AMT.
This is an important advance because an OP description is conceptually
simpler, and physically more intuitive, than the standard sigma model
description of the AMT.\cite{R} Renormalization of this OP field theory then
generates the expected RF structure, which for unknown reasons is not
present in the bare theory. The upper critical dimension $d_c^{+}$ is found
to be $d_c^{+}=6.$ For $d>6$, mean-field theory gives the exact critical
behavior, and for $d<6$, an $\varepsilon =6-d$ expansion for the critical
exponents can be obtained. One of the important results is that hyperscaling
is violated at the AMT due to a dangerous irrelevant variable. As a
consequence, Wegner's scaling law near the metal-insulation transition is
modified.

In Section\ III we reviewed two distinct scaling scenarios for the AMT. The
first one was a conventional scaling theory, in the presence of a dangerous
irrelevant variable. The second one introduced the idea that activated
scaling might be relevant near the AMT. Physically, one of the main results
in this second approach is that static or zero-temperature scaling is
expected to set in only at exponentially low temperatures, and that for
practical purposes it is inaccessible close to the AMT.

Electron-electron interactions are necessary in order
for the AMT discussed here to
exist, since for noninteracting electrons one has an Anderson transition
with an uncritical DOS.\cite{WegnerDOS} This point is correctly reflected by
the theory since $f_n$ in Eq.\ (\ref{eq:2.8a}) vanishes for noninteracting
systems, so that the critical point discussed here is never reached: For $%
K_{s,t}\rightarrow 0$ the critical disorder for the AMT increases without
bound, $G_c\rightarrow \infty $. This suggests a number of distinct phase
transition scenarios. The simplest one is that for sufficiently small
interaction constants, or large $G_c$, the AMT discussed here gets preempted
by some other transition, such as a pure Anderson transition. This scenario
is particularly likely if $K_t=0$, and if the electron-electron interactions
are short ranged, since in this case $K_s$ is irrelevant near the Anderson
transition FP, at least near $d=2$. For this case a likely phase diagram is
shown in Fig.\ \ref{fig:1}. A different possibility is that in the above
picture the Anderson transition is replaced by an AMT of a different type
than the one discussed here, possibly one that is related to the transition
studied near $d=2$ for the case when either $K_s$ and $K_t$ are nonzero, or
the electron-electron interaction is of long range.\cite{R}

\begin{figure}
\epsfxsize=8.25cm
\epsfysize=6.6cm
\epsffile{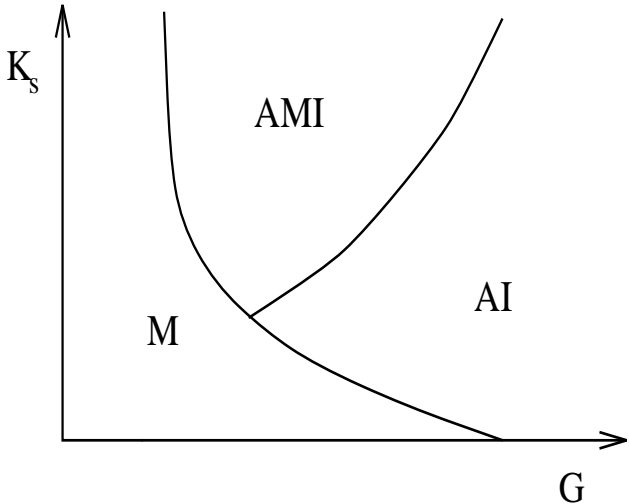}
\vskip 0.5cm
\caption{Schematic phase diagram in the disorder ($G$) - interaction ($K_s$)
plane proposed for a system with $K_t=0$ and a short-ranged $K_s$. $M$, $AI$%
, and $AMI$ denote a metal phase, an Anderson insulator, and an
Anderson-Mott insulator, respectively. The transition from $M$ to $AI$ is an
Anderson transition, while the one from $M$ to $AMI$ is an AMT.}
\label{fig:1}
\end{figure}

In Section III.B we suggested that the AMT is a quantum glass 
transition.\cite{belitz52}
Following this notion, our chief results are as follows: (1)
The specific heat and spin susceptibilities are singular as $T\rightarrow 0$
even in the metallic phase. These results are consistent with existing
experiments, and the theory given here provides an alternative to the
previous exploration in terms of noninteracting local moments. (2) The DOS
is the order parameter for the quantum glass transition. At criticality it
is predicted to vanish logarithmically with temperature. (3) The electrical
conductivity $\tilde{\sigma}$ is so broadly distributed that it is not a
self-averaging quantity, but log $\tilde{\sigma}$ is both self-averaging and
a scaling quantity. This result may be relevant to explain the
sample-to-sample fluctuations in the conductivity that are observed in Si:P
at low temperature near the AMT.

\acknowledgments
This work was supported by the NSF under grant numbers DMR-96-32978 and
DMR-95-10185.

\vfill\eject

\end{mathletters}

\end{document}